\documentclass[12pt,epsf,qsymbols]{article}
\usepackage{tabularx}
\usepackage{array}
\usepackage{graphics}
\usepackage{graphicx}
\usepackage{psfrag}
\usepackage{epsfig}
\usepackage{amsmath,mathbbol}
\usepackage{amssymb}
\usepackage{slashed}
\usepackage{ulem}
\usepackage{setspace}
\usepackage{rotating}
\usepackage{colortbl}
\usepackage{tabularx}
\usepackage{longtable}
\usepackage{lineno}
\makeatletter
\usepackage{textcomp}
\usepackage[usenames,dvipsnames]{xcolor}
\usepackage{relsize}
\usepackage{bbold}

\usepackage{verbatim}
\usepackage{cite}

\setlongtables

\newcommand{\bea}{\begin{eqnarray}}
\newcommand{\eea}{\end{eqnarray}}
\newcommand{\beq}{\begin{equation}}
\newcommand{\eeq}{\end{equation}}
\newcommand{\ec}{\end{center}}
\newcommand{\bc}{\begin{center}}

\newcommand{\gev}{{\rm GeV}}

\newcommand{\pdir}{p\kern -5.2pt\raise 0.2ex\hbox {/}}

\newcommand{\vdir}{v\kern -5.75pt\raise 0.15ex\hbox {/}}
\newcommand{\kdir}{k\kern -5.75pt\raise 0.15ex\hbox {/}}
\newcommand{\epsdir}{\epsilon\kern -5.0pt\raise 0.15ex\hbox {/}}
\newcommand{\bvdir}{\bar{v}\kern -5.75pt\raise 0.15ex\hbox {/}}
\newcommand{\Ddir}{D\kern -7.75pt\raise 0.20ex\hbox {/}}
\newcommand{\Adir}{A\kern -7.75pt\raise 0.20ex\hbox {/}}
\newcommand{\ldir}{l\kern -5.0pt\raise 0.2ex\hbox{/}}
\newcommand{\varepsdir}{\varepsilon\kern -5.5pt\raise 0.15ex\hbox{/}}

\newcommand{\nn}{\nonumber}
\newcommand{\non}[0]{\nonumber \\}
\newcommand{\bee}[0]{\begin{eqnarray}}
\newcommand{\eee}[0]{\end{eqnarray}}
\makeatother

\begin{document}
\thispagestyle{empty} 
\begin{flushright}
\begin{tabular}{l}
{\tt \footnotesize \begin{tabular}{r} LPT 15-10\\ SISSA 09/2015/FISI \end{tabular}}\\
\end{tabular}
\end{flushright}
\begin{center}
\vskip 2.8cm\par
{\par\centering \textbf{\LARGE  
\Large \bf Lepton flavor violating  decays of }}\\
\vskip .35cm\par
{\par\centering \textbf{\LARGE  
\Large \bf vector  quarkonia and of the $Z$ boson}}\\
\vskip 1.05cm\par
{\scalebox{.82}{\par\centering \large  
\sc A.~Abada$^{a}$, D.~Be\v{c}irevi\'c$^{a}$, M.~Lucente$^{a,b}$ and
  O.~Sumensari$^{a,c}$ }
{\par\centering \vskip 0.65 cm\par}
{\sl 
$^a$~Laboratoire de Physique Th\'eorique (B\^at.~210)\\
Universit\'e Paris Sud and CNRS (UMR 8627), F-91405 Orsay-Cedex, France.}\\
{\par\centering \vskip 0.25 cm\par}
{\sl 
$^b$~Scuola Internazionale Superiore di Studi Avanzati,\\
via Bonomea 265, 34136 Trieste, Italy.}\\
{\par\centering \vskip 0.25 cm\par}
{\sl 
$^c$~Instituto de F\'\i sica, Universidade de S\~ ao Paulo,\\ 
C. P. 66.318, 05315-970 S\~ao Paulo, Brazil.}\\

{\vskip 1.65cm\par}}
\end{center}

\vskip 0.55cm
\begin{abstract}
We address the impact of sterile fermions on the lepton flavor violating decays of quarkonia as well as of the $Z$ boson. 
 We compute the relevant Wilson coefficients and show that the ${\rm B}(V\to\ell_\alpha\ell_\beta)$, where $V=\phi,\psi^{(n)}$,  $\Upsilon^{(n)},Z$ can be significantly enhanced 
 in the case of large sterile fermion masses and a non-negligible active-sterile mixing. We illustrate that feature in a specific minimal realization of the inverse seesaw mechanism, known as $(2,3)$-ISS, and in an effective 
 model in which the presence of nonstandard sterile fermions is parameterized by means of one heavy sterile (Majorana) neutrino. 
\end{abstract}
\vskip 2.6cm
{\small PACS: 14.60.St, 14.60.Pq, 13.20.Gd, 13.38.Dg.} 
\newpage
\setcounter{page}{1}
\setcounter{footnote}{0}
\setcounter{equation}{0}
\noindent

\renewcommand{\thefootnote}{\arabic{footnote}}

\setcounter{footnote}{0}
\section{\label{sec-0}Introduction}
So far no signal of new physics has been observed but its search is important in order to understand how to enlarge the Standard Model (SM) to solve 
both the hierarchy and the flavor problems. One of the most significant observations requiring us to go beyond the Standard Model  
is the assessment that neutrinos are massive and that they mix~\cite{Gonzalez-Garcia:2014bfa}. 
Possible SM extensions aiming at incorporating massive neutrinos give rise to interesting collider signatures and open the door 
to new phenomena such as lepton flavor violating (LFV) decays.

Currently, the search for manifestations of LFV constitutes a goal of 
several experimental facilities dedicated to rare lepton decays, such as $\ell_\alpha \to \ell_\beta \gamma$ and $\ell\to \ell_\alpha\ell_\beta \ell_\gamma$, and to the neutrinoless
$\mu-e$ conversion in muonic atoms. One of the most stringent bounds from these searches is the one derived by the MEG Collaboration, 
$\text{B}(\mu \to e \gamma) < 5.7 \times 10^{-13}$~\cite{Adam:2013mnn}, which is expected to be improved to a planned sensitivity of $6 \times 10^{-14}$~\cite{Baldini:2013ke}. 
Moreover, the bound  $\text{B}(\mu \to eee)
< 1.0 \times 10^{-12}$, set by the SINDRUM experiment~\cite{Bellgardt:1987du}, is expected to be improved by the Mu3e experiment where a sensitivity $\sim 10^{-16}$ is planned~\cite{Blondel:2013ia}. 
Limits on the $\tau$ radiative decays~\cite{Aubert:2009ag} and the three-body
decays of $\tau$~\cite{Hayasaka:2010np,Aushev:2010bq} appear to be less
stringent right now, but are likely to be improved at Belle II~\cite{Aushev:2010bq}, where the search for LFV decays of the $B$-meson will be made too~\cite{Bevan:2014iga}. 
The most promising  developments regarding LFV  are those related to the $\mu-e$ conversion in nuclei. The present bound for the  $\mu^- \mathrm{Ti} \rightarrow e^- \mathrm{Ti}$ conversion rate is  $4.3\times
10^{-12}$~\cite{Dohmen:1993mp}, and the planned sensitivity is $\sim10^{-18}$~\cite{Alekou:2013eta}. Similar is the case for gold and aluminum~\cite{Bertl:2006up,Kuno:2013mha}.

Searches for  LFV are also conducted in high-energy experiments and a first bound on the Higgs boson LFV decay $h \to \mu \tau$ has been reported by the CMS Collaboration
 \cite{CMS:2014hha}. The LHCb Collaboration, instead, reported the bound $\text{B}(\tau \to 3 \, \mu)< 8.0 \times 10^{-8}$~\cite{Aaij:2013fia}, which is likely to be improved in the near future~\cite{PDG}. 
 Notice also that they already improved the bounds on $\text{B}(B_{(s)} \to e \mu)$ by an order of magnitude~\cite{Aaij:2013cby}. \\

In this work we will focus on the indirect probes of new physics through the LFV processes of  neutral vector bosons, namely 
$V\to \ell_\alpha\ell_\beta$, with $\ell_{\alpha,\beta}\in \{e,\mu, \tau\}$, and $V\in \{ \phi, \psi^{(n)}, \Upsilon^{(n)}, Z\}$, where $\psi^{(n)}$ stands for $J/\psi$ and its radial excitations, and similarly for $\Upsilon^{(n)}$.  
Most of the research in this direction reported so far is related to the $Z\to \ell_\alpha\ell_\beta$ decay modes. 
More specifically, the experimental bounds,  obtained at LEP are found to be   $\mathrm{B}(Z\to e^\mp \mu^\pm) < 1.7 \times 10^{-6}$~\cite{Abreu:1996mj}, $\text{B}(Z\to \mu^{\mp}\tau^{\pm})< 1.2\times 10^{-5}$~\cite{Akers:1995gz,Abreu:1996mj}, 
and $\text{B}(Z\to e^{\mp}\tau^{\pm})< 9.8 \times 10^{-6}$~\cite{Akers:1995gz,Adriani:1993sy}. 
One of these bounds has been improved at LHC, namely $\text{B}(Z\to e^{\mp}\mu^{\pm})< 7.5\times 10^{-7}$~\cite{Aad:2014bca}. On the theory side, the $Z$ decays have been analyzed in the extensions of the SM involving 
additional massive and sterile neutrinos that could mix with the standard (active) ones and thus give rise to the LFV decay rates~\cite{Mann:1983dv,Ilakovac:1994kj,Illana:1999ww}. A similar approach has been also adopted in Ref.~\cite{Abada:2014cca}, in the perspective of 
a Tera-$Z$ factory FCC-ee~\cite{FCC-WG} for which a targeted sensitivity is expected to be $\text{B}(Z\to e^{\mp}\mu^{\pm}) \,\sim \, 10^{-13}$~\cite{Blondel:2014bra}.

Lepton flavor conserving decays of quarkonia have been measured to a high accuracy which can actually be used to fix the hadronic parameters (decay constants). Otherwise, one can use the results of 
numerical simulations of QCD on the lattice, which are nowadays accurate as well~\cite{c-latt1,c-latt2,Lepage,RL}. The experimentally established bounds for the simplest LFV decays of quarkonia are~\cite{PDG}:
\bea
&&\text{B}(\phi \to e \mu) < 2.0\times 10^{-6},   \qquad  \text{\cite{Achasov:2009en}}\nn \\
&&\hfill \nn\\
&&\text{B}(J/\psi \to e \mu) < 1.6\times 10^{-7}, \qquad \text{B}(J/\psi \to e \tau) < 8.3\times 10^{-6},  \nn \\
 &&\text{B}(J/\psi \to  \mu \tau) < 2.0\times 10^{-6} ,  \qquad  \text{\cite{Ablikim:2013qtm,Ablikim:2004nn}} \nn \\
&&\hfill \nn\\
&&\text{B}(\Upsilon \to  \mu \tau ) < 6.0\times 10^{-7} ,   \qquad \text{\cite{Love:2008ys}}   \nn \\
&&\hfill \nn\\
&& \text{B}(\Upsilon(2S) \to e \tau) < 8.3\times 10^{-6},\qquad  \text{B}(\Upsilon(2S) \to  \mu \tau) < 2.0\times 10^{-6}, \qquad \text{ \cite{Lees:2010jk}} \nn \\
&& \text{B}(\Upsilon(3S) \to e \tau) < 4.2\times 10^{-6},\qquad  \text{B}(\Upsilon(3S) \to  \mu \tau) < 3.1\times 10^{-6} , \qquad \text{ \cite{Lees:2010jk}}  \nn 
\eea
where each mode is to be understood as $\text{B}(V \to \ell_\alpha\ell_\beta) = \text{B}(V \to \ell_\alpha^+ \ell_\beta^-)+ \text{B}(V \to \ell_\alpha^- \ell_\beta^+)$.

Despite the appreciable experimental work on the latter observables, only a few theoretical studies have been carried out so far.  
The authors of Ref.~\cite{Nussinov:2000nm} applied a vector meson dominance approximation to $\mu \to 3 e$ and expressed the width of the latter process, 
$\Gamma(\mu \to ee e)=\Gamma(\mu\to V e )\Gamma(V\to ee)$. Since the values of $\Gamma(V\to ee)$ are very well known experimentally~\cite{PDG}, the experimental 
bound on $\Gamma(\mu \to 3 e)$ is then used to obtain an upper bound on the phenomenological coupling $g_{V\mu e}$, which is then converted to an upper bound on $\Gamma( V \to \mu e )$.  
A similar approach has been used in Ref.~\cite{Gutsche:2009vp} where instead of $\mu \to eee$, the authors considered the $\mu-e$ conversion in nuclei ($N$), which they described in terms 
of a product of couplings $g_{V\mu e}$ and $g_{V N N}$. 
The latter could be extracted from the experimentally measured $\Gamma(V \to p\bar p)$, and with that knowledge the experimental upper bound on  $\text{R}(\mu \text{Ti}\to e\text{Ti})$ results in an upper bound 
on $\Gamma( V \to \mu e )$. 
A more dynamical approach in modeling the $V \to \ell_\alpha\ell_\beta$ processes has been made in a supersymmetric extension of the SM with type I seesaw~\cite{Sun:2012yq}. 

Sterile fermions were proposed in various neutrino mass generation mechanisms, but the interest in their properties was further motivated by the reactor/accelerator 
anomalies~\cite{reactor:I,Aguilar:2001ty,miniboone:I,gallium:I},  a possibility to offer a warm dark matter 
candidate~\cite{Dodelson:1993je,Abazajian:2001nj,nu_WDM}, and by indications from the large scale structure
formation~\cite{general_structure,Kusenko:2009up,Abazajian:2012ys}.

Incorporating neutrino oscillations (masses and mixing~\cite{Gonzalez-Garcia:2014bfa}) into the  SM implies that the  charged current 
is modified to 
\begin{equation}\label{eq:cc-lag1}
- \mathcal{L}_\text{cc} = \frac{g}{\sqrt{2}} U^{\alpha i} 
\bar{\ell}_\alpha \gamma^\mu P_L \nu_i  W_\mu^- + \, \text{c.c.}\,,
\end{equation}
$U$ being the leptonic mixing matrix,  $\alpha$ the flavor of a charged lepton, and 
$i = 1, \dots, n_\nu$ denotes a physical neutrino state. If one assumes that  only three massive neutrinos are present,  
the matrix $U$ corresponds to the unitary Pontecorvo-Maki-Nakagawa-Sakata (PMNS) matrix. In that situation the GIM mechanism makes the decay 
rates B($V \to \ell_\alpha^\mp \ell_\beta^\pm$) completely negligible, $\lesssim10^{-50}$. That feature, however, can be drastically 
changed in the presence of a non-negligible mixing with heavy sterile fermions.
In what follows we will consider such situations, derive analytical expressions for B($V \to \ell_\alpha\ell_\beta$), and discuss 
a specific realization of the inverse seesaw mechanism, known as (2,3)-ISS~\cite{Abada:2014vea}. We will also discuss a 
simplified model in which the effect of the heavy sterile neutrinos is described by one effective sterile neutrino state with non-negligible mixing with active neutrinos.~\footnote{In this work, 
due to the tension between the most recent Planck results on extra light neutrinos (relics) and the reactor/accelerator  anomalies, we will consider the effect of (heavier) sterile neutrinos not 
contributing as light relativistic degrees of freedom~\cite{Ade:2013zuv}. We will require our models to be compatible with current experimental data and constraints and to fulfill the so-called perturbative 
unitary condition which puts a strong constraint on the models for the very heavy sterile fermion(s)~\cite{Chanowitz:1978mv}.} 
Despite several differences, our approach is similar to the one discussed in Ref.~\cite{Ilakovac:1999md}, where the SM has been extended by new, heavy, Dirac neutrinos, singlets under $SU(2)\times U(1)$, and 
applied to a number of low energy decay processes. Our sterile neutrinos are Majorana and we apply the approach to the leptonic decays of quarkonia for the first time. 

The remainder of this paper is organized as follows: In Sec.~\ref{sec:obs} we formulate the problem in terms of a low energy effective theory of a larger theory which contains heavy sterile neutrinos, 
we derive expression for B($V \to \ell_\alpha  \ell_\beta $) and compute the Wilson coefficients.  In Sec.~\ref{sec:ext} we briefly describe the specific models with sterile neutrinos which are used in this paper to produce our results presented 
in Sec.~\ref{sec:results}. We finally conclude in Sec.~\ref{sec:concl}.

\section{LFV decay of Quarkonia - Effective Theory\label{sec:obs}}

In this section we formulate a low energy effective theory of the LFV decays of quarkonia of type $V\to \ell_\alpha^\pm\ell_\beta^\mp$, and express the decay amplitude in terms of the quarkonium decay constants and the corresponding Wilson coefficients. 
The latter are then computed in the extensions of the SM which include the heavy sterile neutrinos.  We also derive the expression relevant to  $\Gamma(Z\to \ell_\alpha^\pm\ell_\beta^\mp)$.

\subsection{Effective Hamiltonian} 

Keeping in mind the fact that we are extending the SM by adding sterile fermions, without touching the gauge sector of the theory, the decays of vector quarkonia, 
$V (q) \to \ell_\alpha^\pm (p)\ \ell_\beta^\mp (q-p)$, can only occur through the photon and 
the $Z$-boson exchange at tree level. In the lepton flavor conserving processes the $Z$-exchange terms are very small with respect to those arising from the electromagnetic interaction and are usually neglected. The generic effective Hamiltonian can be written as
\begin{equation}\label{Heff:V}
	\mathcal{H}_{\text{eff}} = {\cal Q}_Q \frac{e^2 g^2}{2 m_V^2} \, \bar{Q} \gamma_\mu Q\, \cdot\, \bar{\ell}_\alpha \left[C_{VL} \gamma^\mu P_L +C_{VR} \gamma^\mu P_R +  \frac{p^\mu}{m_W} (C_R P_R + C_L P_L)\right] \ell_\beta,
\end{equation}
where $ {\cal Q}_Q$ is the electric charge of the quark $Q$, $m_V$ is the mass of quarkonium $V$ which is dominated by the valence quark configuration $\bar Q Q$,~\footnote{We remind the reader that the ground vector meson $\bar s s$, $\bar c c$, $\bar b b$ states are $\phi$, $J/\psi$, $\Upsilon$, respectively, and the corresponding charges are  $ {\cal Q}_{s,b}=-1/3$ and  $ {\cal Q}_c=2/3$.}  $C_{VL,VR,L,R}$ are the Wilson coefficients,  $p$ is the momentum of one of the outgoing leptons, and $P_{L/R}= \frac{1}{2}(1 \mp \gamma_5)$. Contributions to the scalar (left and right) terms are suppressed by $m_{\alpha,\beta}/m_W$, where $m_{\alpha,\beta}$ are the charged lepton masses. In this section we will keep such terms so that our expressions can be useful to approaches in which the scalar bosons are taken  in consideration. For our phenomenological discussion, however, it is worth emphasizing that $C_{L,R,VR}\ll C_{VL}$. 
\begin{figure}[h!]
\begin{center}
\includegraphics[scale=.38]{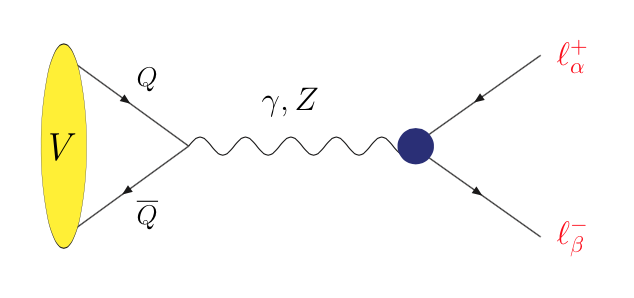}~\includegraphics[scale=.38]{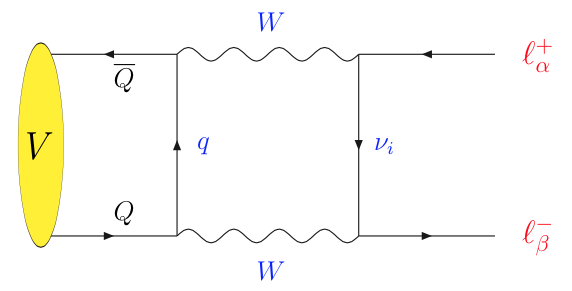}
\caption{{\footnotesize 
Diagrams contributing the LFV decay of quarkonia $V\to \ell_\alpha \ell_\beta$. The blob in the first diagram is related to the penguin loop that generates the LFV, and the box diagram is particularly important to be included in the case of 
$\Upsilon^{(n)}\to   \ell_\alpha \ell_\beta$ because of $V_{tb}\simeq 1$ and of the top quark mass, making the box diagram contribution to the Wilson coefficient significant.
 }}
\label{fig:1}
\end{center}
\end{figure}

Without entering the details of calculation it is easy to verify that the only relevant diagrams are those shown in Fig.~\ref{fig:1},
and therefore the structure of  the Wilson coefficients $C_i$ reads,
\begin{equation}
C_{i}=C^\gamma_{i} +  C^Z_{i} \frac{1}{\sin^2\theta_W\cos^2\theta_W}\frac{m_V^2}{m_V^2-m_Z^2} \frac{g_{V}^{\cal Q}}{{\cal Q}_Q}+C^{\mathrm{Box}}_{i} |V_{Qq}|^2\frac{1}{\sin^2\theta_W} \frac{m_V^2}{m_W^2} \frac{1}{{\cal Q}_Q},
\end{equation}
where $C^{\gamma,Z}_i$ are  the contributions arising from the photon and the $Z$-boson exchange, while $C_i^{\mathrm{Box}}$ comes from the box diagram that  involves the Cabibbo--Kobayashi-Maskawa 
coupling $V_{Qq}$.~\footnote{The box diagram contribution to $V \to \ell_\alpha\ell_\beta$ in the case of $V=\Upsilon$ is dominated by the top quark ($|V_{tb}|\simeq 1$); for $V= \psi$ it is negligible because 
the contribution of the $b$ quark is Cabibbo suppressed ($|V_{cb}|\simeq 0.004$) while the Cabibbo allowed one ($|V_{cs}|\simeq 0.99$) is suppressed by the strange quark mass; for $V=\phi$, the 
contributions of the charm and top quarks are comparable but overall smaller than in the $\Upsilon \to \ell_\alpha\ell_\beta$ case. 
} In the above expressions $g_V^Q=\frac{1}{2}I_3^Q-{\cal Q}_Q \sin^2\theta_W$.
The blob in the diagram shown in Fig.~\ref{fig:1} stands for the lepton loop diagrams that may contain one or two neutrino states and which, in the extensions of the SM involving a heavy neutrino sector, will give rise to the LFV decay  due to the effect of mixing which is parametrized by the matrix $U$ [see Eq.~(\ref{eq:cc-lag1})]. 
Separate contributions coming from different diagrams can be further reduced by factoring out the neutrino mixing matrix elements, namely
\begin{equation}
C_{i}^{\gamma,\mathrm{Box}}=\sum_{k=1}^{n_\nu} U_{\beta k} U_{\alpha k}^* C^{\gamma,\mathrm{Box}; k}_{i},  \quad \text{and} \quad C_{i}^Z=\sum_{k=1}^{n_\nu} U_{\beta k} U_{\alpha k}^* C^{Z, k}_{i}+\sum_{k=1}^{n_\nu}\sum_{j=1}^{n_\nu} U_{\beta k} U_{\alpha j}^* C^{Z, kj}_{i},
\end{equation}
where we see that the term involving two neutrino eigenstates appears only in the $Z$ coefficient because it is related to the vertex $Z \nu_k\nu_j$.
It is worth emphasizing that the tensor structure in Eq.~(\ref{Heff:V}) 
can be easily obtained from the coefficients $C_{L,R}$  by applying the Gordon identity. 
Such contributions are $1/m_W$ suppressed, and thus completely negligible,  which is why we do not give explicit expressions for these coefficients.

Using the effective Hamiltonian (\ref{Heff:V}) and parameterizing the hadronic matrix as 
\bea
\langle 0 \vert  \bar Q \gamma_\mu Q \vert V(q,\sigma) \rangle = f_V m_V \varepsilon_\mu^\sigma\,, 
\eea
where $f_V$ is the decay constant of a quarkonium $V$ with momentum $q$ and in a polarization state $\sigma$, we can write the decay rate as,  
\begin{align}
\Gamma(V\to{\ell}_\alpha^- \ell_\beta^+ ) = \frac{8 \pi {\cal Q}_Q^2 \alpha^2}{3 m_V^3}  G_F^2 m_W^4 \left(\frac{f_V}{m_V}\right)^2  {\lambda^{1/2}(m_V^2,m_\alpha^2,m_\beta^2)} \phi_C,
\end{align}
with 
\begin{equation}
\lambda(a^2,b^2,c^2)=[a^2-(b-c)^2][a^2-(b+c)^2],\label{lambdadef}\end{equation} and 
\begin{align}
\phi_C=\left(-g^{\mu \nu}+\frac{q^\mu q^\nu}{m_V^2} \right) \mathrm{tr} \Big[(\slashed{q}-\slashed{p}+m_\beta) \cdot (C_{VL} \gamma^\mu P_L + C_{VR} \gamma^\mu P_R+C_L \frac{p^\mu}{m_W}P_L+C_R \frac{p^\mu}{m_W}P_R) \nonumber \\
\cdot (\slashed{p}-m_\alpha)\cdot (C_{VL}^* \gamma^\nu P_L + C_{VR}^* \gamma^\nu P_R+C_L^* \frac{p^\nu}{m_W}P_R+C_R^* \frac{p^\nu}{m_W}P_L)\Big],
\end{align}
which gives
\begin{align}
\phi_C=\frac{1}{4 m_V^2 m_W^2} &\Big{\lbrace}\lambda(m_V^2,m_\alpha^2,m_\beta^2)\Big[(m_V^2-m_\alpha^2-m_\beta^2)(|C_L|^2+|C_R|^2)-4\mathrm{Re} (C_L^* C_R) m_\alpha m_\beta \nn \\ 
&+4 m_W\mathrm{Re}(C_L^* (C_{VL} m_\beta+C_{VR} m_\alpha)+C_R^* (C_{VL} m_\alpha+ C_{VR} m_\beta))\Big] \nn\\
&+ 4 m_W^2 (|C_{VL}|^2+|C_{VR}|^2) \Big[2 m_V^4 - m_V^2(m_\alpha^2+m_\beta^2)- (m_\alpha^2-m_\beta^2)^2 \Big]\\
&+ 48 m_W^2 m_V^2 m_\alpha m_\beta   \mathrm{Re} (C_{VL}^* C_{VR}) \nn 
 \Big{\rbrace}.
\end{align}
As we mentioned above, we consider in our framework $C_{VL}\gg C_{VR,R,L}$, and therefore we can write 
\begin{align}
\Gamma (V\to \ell_\alpha^\pm\ell_\beta^\mp)=\frac{32\pi {\cal Q}_Q^2 \alpha^2}{3 m_V^3}  f_V^2 G_F^2 m_W^4   |C_{VL}|^2 
 \lambda^{1/2}(m_V^2,m_\alpha^2,m_\beta^2)\Big{[}1-  \frac{(m_\alpha^2+m_\beta^2)}{2 m_V^2}-\frac{(m_\alpha^2-m_\beta^2)^2}{2 m_V^4}\Big{]},
\end{align}
where $\lambda(a^2,b^2,c^2)$ is given in Eq.~(\ref{lambdadef}). In this last expression we also used $\Gamma (V\to \ell_\alpha^\pm\ell_\beta^\mp) = \Gamma (V\to \ell_\alpha^+\ell_\beta^-)+\Gamma (V\to \ell_\alpha^-\ell_\beta^+)$.

Besides quarkonia we will also revisit the issue of adding extra species of sterile neutrinos to the decay of $Z\to \ell_\alpha^\pm\ell_\beta^\mp$. In that case the effective Hamiltonian can be written as  
\begin{equation}
\mathcal{H}^Z_{\mathrm{eff}}= \frac{g^3}{2 \cos\theta_W} \bar{\ell}_\alpha \Big{[} D_{VL}\gamma^\mu P_L + D_{VR}\gamma^\mu P_R +D_{L}P_L+D_{R}P_R\Big{]}\ell_\beta Z^\mu, 
\end{equation}
where the Wilson coefficients are now denoted by $D_i$ and take the form  
\begin{equation}
	D_{i}=\sum_{k=1}^{n_\nu} U_{\beta k} U_{\alpha k}^* C^{Z, k}_{i}+\sum_{k =1}^{n_\nu}\sum_{j=1}^{n_\nu} U_{\beta k} U_{\alpha j}^* C^{Z, kj}_{i}\,.
\end{equation}
The decay rate in the similar limit, $D_{VL}\gg D_{VR,R,L}$, reads 
\begin{align}
\label{gammaz}
\Gamma(Z\to{\ell}_\alpha^- \ell_\beta^+ ) = \frac{8 \sqrt{2}}{3 \pi m_Z} \frac{G_F^3 m_W^6}{\cos^2\theta_W} |D_{VL}|^2 \lambda^{1/2}(m_Z^2,m_\alpha^2,m_\beta^2) \left[1-\frac{(m_\alpha^2+m_\beta^2)}{2 m_Z^2}-\frac{(m_\alpha^2-m_\beta^2)^2}{2 m_Z^4}\right].
\end{align}

\subsection{Wilson coefficients}
Concerning the computation of the Wilson coefficients we stress again that our results are obtained in a theory in which the Standard Model is extended to include extra species of sterile fermions, without changing the gauge sector. 
The origin  of the leptonic mixing matrix $U$ is model dependent and in order to be able to do a phenomenological analysis,  we will have to adopt a specific model which will be discussed in the next section. 

\begin{figure}[t!]
\hspace*{-11mm}\begin{tabular}{cccccc}
\includegraphics[scale=.28]{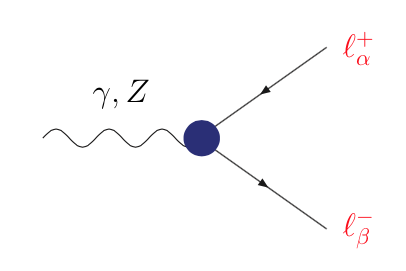}&$\raise 7.4ex \hbox{=}$&\includegraphics[scale=.28]{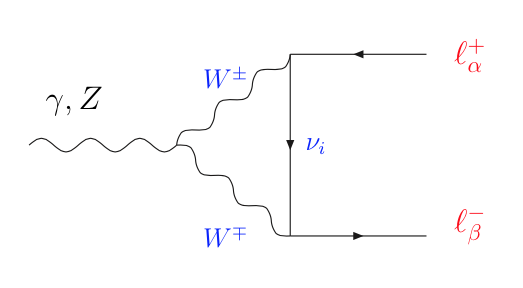}&$\raise 7.4ex \hbox{+}$&\includegraphics[scale=.28]{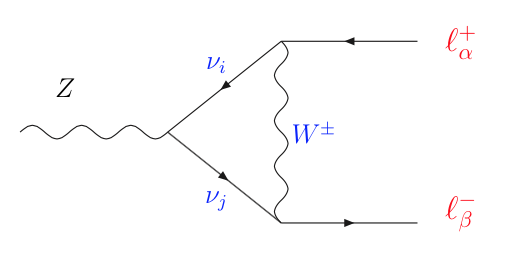}&$\raise 7.4ex \hbox{+~\dots}$
  \end{tabular}
  \caption{{\footnotesize 
Vertex diagrams contributing the LFV decays.
 }}
\label{fig:2}
\end{figure}

The blob in the diagram shown in Fig.~\ref{fig:1} stands for a series of diagrams such as those displayed in Fig.~\ref{fig:2}. All of them, including the box diagram in Fig.~\ref{fig:1}, have been computed in the Feynman gauge and the results are collected in Appendix~A.
Here we focus on the most important contributions in the case of large masses of sterile (Majorana) neutrinos. Contributions to the Wilson coefficients coming from vertex diagrams can be divided into two pieces: those involving only one neutrino in the loop, $C_{VL}^{Z,\gamma}(x_i)$, where $x_i = m_i^2/m_W^2$, and those with two neutrinos in the loop, $C_{VL}^{Z }(x_i ,x_j)$. In the limit of large values of $x_{i,j} \gg 1$, we find the following behavior 
\begin{align}\label{eq:Wils}
C_{VL}^{Z}(x_i)  \stackrel{x_{i}\gg 1}{\xrightarrow{\hspace*{9mm}}} &\,  \frac{5}{32\pi^2} \log x_i + \text{\small finite term} + {\cal O}(1/x_i) \sim \log x_i\, , \cr
 C_{VL}^{Z }(x_i ,x_i) \stackrel{x_{i}\gg 1}{\xrightarrow{\hspace*{9mm}}} &\,    \frac{C_{ii}}{64 \pi^2}\left\lbrace  \left( 2 x_i +3 -4 \log x_i\right)+x_i\left(\log x_i-\frac{7}{2}\right)\right\rbrace +\dots \cr  & \ \sim C_{ii}\ x_i \log x_i +  \dots 
\end{align}

\begin{figure}[h!]
\begin{center}
\hspace*{-4mm}\includegraphics[scale=.42]{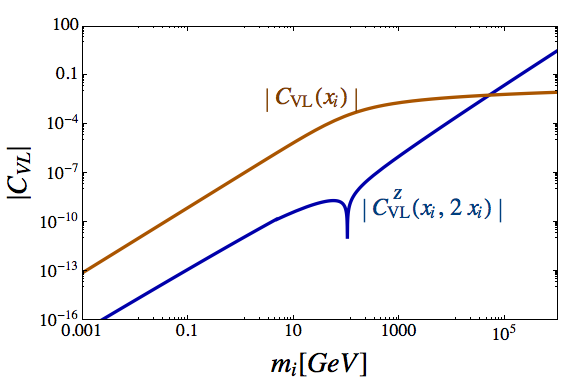}~\includegraphics[scale=.42]{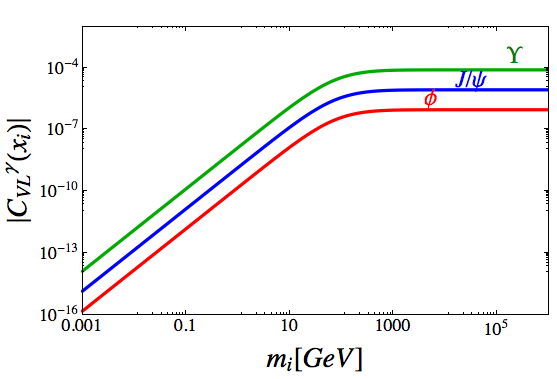}
\caption{{\footnotesize 
In the left panel are shown $C_{VL}(x_i)$ and $C_{VL}^Z(x_i ,x_j)$, for $x_j=2 x_i$,  as functions of $m_i =  m_W \sqrt{x_i}$,  the mass of the heavy sterile neutrino propagating in the loops. For illustration purpose, the couplings $C_{ij}$ were fixed to a common value, $10^{-5}$, and the example corresponds to the $\Upsilon \to \mu\tau$ decay. Right panel:  $C_{VL}^{\gamma}(x_i)$  is plotted as a function of $m_i$ for the case of $V\to e\mu$ in three specific cases $V\in \{\phi, J/\psi,\Upsilon\}$. In both cases the value of functions at $x_{i,j}=0$ have been subtracted away. }}
\label{fig:3}
\end{center}
\end{figure}

To illustrate the relative contribution of the different diagrams we fix the values of the coefficients $C_{ij}\equiv \displaystyle{\sum_{\alpha=e,\mu,\tau}} U_{\alpha i}^\ast U_{\alpha j} = 10^{-5}$, and plot  $|C_{VL}(x_i)-C_{VL}(0)|$ and $|C_{VL}^{Z}(x_i,2\ x_i)-C_{VL}^{Z}(0,0)|$ for the case of $\Upsilon \to \mu\tau$, cf. Fig.~\ref{fig:3}.~\footnote{Due to the unitarity of the mixing matrix $U$, the terms in the Wilson coefficients that do not depend on neutrino masses give a vanishing contribution after summing over all neutrino states. We thus subtract the constant terms in the plots in order to better appreciate the dependence on the neutrino masses. Notice also that $C_{ij}= 10^{-5}$ is in agreement with all constraints discussed in the text when the neutrino masses are below $\mathcal{O}(100)$ TeV.}  We see that only for very large masses the diagrams with two neutrinos in the loop become more important than those with one neutrino state.  We should stress that each contribution to $C_{VL}(x_i)$, i.e.  $C_{VL}^{\rm Box}(x_i)$ and $C_{VL}^{Z}(x_i)$, scales as $\log x_i$  for large values of $x_i$, except for $C_{VL}^{\gamma}(x_i)$ which goes to a constant in the same limit. That can also be seen in Fig.~\ref{fig:3} where in the left panel we show the dependence of the total $C_{VL}(x_i)$ on $x_i$ and in the right panel we show  $C_{VL}^{\gamma}(x_i)$ and its dependence on the mass of the initial decaying meson, $\phi$, $J/\psi$, and $\Upsilon$.  The contribution of sterile neutrinos to the LFV decay of $\Upsilon$ is larger than the one to lighter mesons, since the Wilson coefficients are also proportional to the mass of the initial particle.

 Before closing this section we should reiterate that our  Wilson coefficients have  been computed in the Feynman gauge. Since  all divergencies cancel out,  our results are finite and gauge invariant, as was already  observed in Refs.~\cite{Mann:1983dv,Ilakovac:1994kj,Illana:1999ww,Abada:2014cca}.

\section{SM in the presence of sterile fermions \label{sec:ext} }
With the expressions derived above, we now have to specify a model for lepton mixing (couplings) $U_{\alpha i}$ in  the presence of heavy sterile neutrinos propagating in the loops. 
We opt for a minimal realization of the inverse seesaw mechanism for the generation of neutrino masses, which is nowadays rather well constrained by the available experimental data. 
Furthermore, we will use a parametric model containing one effective sterile neutrino, which essentially mimics the behavior at low energy scales of mechanisms involving heavy sterile fermions.

\subsection{The (2,3)-inverse seesaw realization}
Among many possible realizations of accounting for massive neutrinos, the inverse seesaw mechanism (ISS)~\cite{ISS} 
offers the possibility of accommodating
the smallness of the active neutrino masses for a comparatively  low seesaw scale, but still with natural $\mathcal{O}(1)$
Yukawa couplings, which renders  this scenario phenomenologically appealing. Indeed, depending on their masses and mixing with active neutrinos, the new  states can be
produced in collider and/or low energy experiments, and their
contribution to physical processes can be sizable.
 ISS, embedded in the SM, results in a mass term for neutrinos of the form   
\begin{equation}
-\mathcal{L}_\text{mass}=\frac{1}{2}n_L^T C M n_L+\mbox{h.c.}\ ,
\end{equation}
where $C \equiv i \gamma^2 \gamma^0$ is the charge conjugation matrix and $n_L \equiv {\left(\nu_{L,\alpha},\nu_{R,i}^c,s_j\right)}^T$\!.  Here $\nu_{L,\alpha}$, $\alpha=e,\mu,\tau$ denotes the active (left-handed)  neutrino states of the SM, while $\nu_{R,i}^c$ ($i=1,  \#\nu_R$) and $s_j$ ($j=1, \# s$) are right-handed neutrino fields and  additional fermionic gauge singlets, respectively.  The  neutrino mass matrix $M$ then has the form
\begin{equation}\label{general-iss}
M \equiv
\left(
\begin{array}{ccc}
0 & d & 0 \\
d^T & 0 & n \\
0 & n^T & \mu
\end{array}
\right)\ ,
\end{equation}
where $d, n,\mu$ are complex matrices.~\footnote{It is in general possible to consider also a nonzero value for the central entry of the matrix (\ref{general-iss}), with elements at a mass scale similar to the one of $\mu$. These parameters, however, only affect neutrino masses and mixing at loop level~\cite{Dev:2012sg}, which is why we do not consider them here.}\\
The Dirac mass matrix $d$ arises from the Yukawa couplings to the SM Higgs boson, $\tilde{H} =i  \sigma^2 H$,
\begin{equation}
Y_{\alpha i} \overline{\ell_L^\alpha}\tilde{H }\nu_R^i+\text{H.c., }\,\,\,\,\,\,\ell_L^\alpha=\left(
\begin{array}{c}
\nu_{L}^\alpha \\
e^\alpha_L
\end{array}
\right)\ ,
\end{equation}
while the matrix $\mu$, instead, contains the Majorana mass terms for the sterile fermions $s_j$. 
By assigning a leptonic charge  $L=+1$ to both $\nu_R$ and $s$, one makes sure that the off diagonal terms are lepton number conserving, while $s^T Cs$ violates the lepton number by two units. 
Furthermore, the interesting feature of this seesaw realization is that the entries of  $\mu$ can be made small in order to accommodate for the ${\mathcal{O}}(\text{eV}$) masses of active neutrinos, with large Yukawa couplings. 
This is not in conflict with naturalness since the lepton number is restored in the limit of $\mu \rightarrow 0$.~\footnote{In this work we consider configurations in which the entries in the above matrices fulfill a {\sl naturalness criterion}, $|\mu|\ll |d|<|n|$~\cite{Abada:2014vea}.}

Concerning the additional sterile states $\nu_R$ and $s$, since up to now there is no direct evidence for their existence and because they do not contribute to anomalies, their number is unknown. 
In Ref.~\cite{Abada:2014vea}  it was shown that it is possible to construct several minimal distinct realizations of ISS, each reproducing  the correct neutrino mass spectrum and  satisfying all phenomenological constraints. 
More specifically, it was shown that, depending on the number of additional fields, the neutrino mass spectrum obtained for each ISS realization is characterized by either two or three mass scales, one corresponding to $m_\nu \approx \mu\ d^2/n^2$ (light neutrino masses), one corresponding to the heavy mass eigenstates [the mass scale of the matrix $n$ of Eq.~(\ref{general-iss})], and finally an intermediate scale $\sim\mu$,  only present if $\# s > \#\nu_R$. 
This allows us to identify  two truly minimal ISS realizations that comply with all experimental bounds, namely 
the (2,2)-ISS model, which corresponds to the SM extended by two right-handed (RH) neutrinos and two additional sterile states, leading  to a three-flavor mixing scheme, and 
the (2,3)-ISS realization, where the SM is extended by two RH neutrinos and three sterile states leading to a 3+1-mixing scheme. Interestingly, the lightest sterile neutrino with a mass around eV in the 
(2,3)-ISS  can be used to explain the short baseline (reactor/accelerator) anomaly~\cite{reactor:I,Aguilar:2001ty,miniboone:I,gallium:I} if its mass lies around eV, or to provide a dark matter candidate if the lightest sterile state were in the keV range~\cite{Abada:2014zra}.

\subsection{A model with one effective sterile fermion}

Since the generic idea of obtaining a significant contribution to our observables applies to any model
in which the active neutrinos have sizable mixing with some additional singlet states (sterile fermions),  
we can use an \textit{effective} model with three light active neutrinos plus one extra sterile neutrino.

The introduction of this extra state implies three new active-sterile mixing angles ($ \theta_{14}, \theta_{24}, \theta_{34}$), two extra Dirac $CP$ violating phases ($\delta_{14},\delta_{34}$) and one additional Majorana phase ($\phi_{41}$). 
The lepton mixing matrix is then a product of six rotations times the Majorana phases, namely
\begin{eqnarray} \label{eq:3+1rot}
U &=& R_{34}(\theta_{34},\delta_{34}) \cdot R_{24}(\theta_{24}) \cdot R_{14}(\theta_{14},\delta_{14}) 
\cdot R_{23} \cdot R_{13} \cdot R_{12}  \cdot \rm diag(\phi_{21},\phi_{31},\phi_{41}) \nonumber \\
&=& R_{34}(\theta_{34},\delta_{34}) \cdot R_{24}(\theta_{24}) \cdot R_{14}(\theta_{14},\delta_{14}) 
\cdot U_{\rm PMNS} \cdot \rm diag(\phi_{21},\phi_{31},\phi_{41})\,,
\end{eqnarray} 
where the rotation matrices $R_{34},R_{24},R_{14}$ can be defined as:
\begin{eqnarray} \label{eq:R}
R_{34}\ &=&\ \left( 
\begin{array}{cccc}
1 & 0 & 0 & 0 \\
0 & 1 & 0 & 0 \\
 0 & 0 & \rm cos \theta_{34} &\rm sin \theta_{34} \cdot e^{-i \delta_{34}}\\ 
 0 & 0 & -\rm sin \theta_{34} \cdot e^{i \delta_{34}}& \rm cos \theta_{34} 
\end{array}%
\right)\,, \nonumber \\
R_{24}\ &=&\  \left( 
\begin{array}{cccc}
1 & 0 & 0 & 0 \\
0 & \rm cos \theta_{24}  & 0 & \rm sin \theta_{24}\\ 
0 & 0 & 1 & 0 \\
0 & - \rm sin \theta_{24}& 0 &  \rm cos \theta_{24} %
\end{array}%
\right)\,, \nonumber \\
R_{14}\ &=& \left( 
\begin{array}{cccc}
\rm cos \theta_{14} & 0 & 0 & \rm sin \theta_{14} \cdot e^{-i \delta_{14}} \\
0 & 1 & 0 & 0 \\
0 & 0 & 1 & 0 \\
- \rm sin \theta_{14} \cdot e^{i \delta_{14}} & 0 & 0 &\rm cos \theta_{14} \\
\end{array}%
\right) \,.
\end{eqnarray} 

In the framework of the SM extended by sterile fermion states, which
have a nonvanishing mixing with active neutrinos, 
the Lagrangian describing the leptonic charged currents becomes
\begin{equation}\label{eq:cc-lag}
- \mathcal{L}_\text{cc} = \frac{g}{\sqrt{2}} U^{\alpha i} 
\bar{\ell}_\alpha \gamma^\mu P_L \nu_i  W_\mu^- + \, \text{c.c.}\,,
\end{equation}
where $i = 1, \dots, n_\nu$ denotes the physical neutrino states, 
and $\alpha = e, \mu, \tau$ are the flavors of the charged leptons. 
In the case of the SM with three neutrino generations,  $U$ is the  PMNS matrix, while in the case of $n_\nu \geq 4$, the 
$3\times 3$ submatrix ($\widetilde U_\text{PMNS}$) is not unitary anymore and one can parameterize it as
\begin{equation}\label{eq:U:eta:PMNS2}
U_\text{PMNS} \, \to \, \widetilde U_\text{PMNS} \, = \,(\mathbb{1} -\widetilde \eta)\, 
U_\text{PMNS}\,,
\end{equation}
where $\widetilde \eta$ is a matrix that accounts for the deviation of $\widetilde
U_\text{PMNS}$ from unitarity~\cite{Schechter:1980gr,Gronau}, due to the presence of extra fermion states.
Many observables are sensitive to the 
active-sterile mixing and their current experimental values can be used to constrain the $\widetilde \eta$ matrix~\cite{Antusch:2014woa}. 

In order to express the deviation from unitarity in terms of a single parameter, we define
\bea
\eta = 1- |\det \widetilde U_\text{PMNS}|  \,,
\eea
which, in the case of the extension of the SM by only one sterile fermion and in terms of the mixing angles defined above, reads
\bea
\eta = 1 - \vert \cos\theta_{14} \cos\theta_{24}\cos\theta_{34}\vert \,.
\eea
\section{Results and discussion}\label{sec:results}

In this section we present and discuss our results.

Since the Wilson coefficients of the processes discussed here are proportional to the mass of the decaying particle, it is quite obvious that the most significant enhancement of B($V \to \ell_\alpha  \ell_\beta$) will occur for $V=\Upsilon$ and its radial excitations. 
For this reason we will present plots of our results for this decay channel. Plots for other channels are completely similar which is why we do not display them. Before we discuss the impact of the active-sterile neutrino mixing on the LFV decay rates further, we first specify the constraints on parameters of
both of our models.  
 
\begin{figure} 
\begin{center}
\hspace*{-14mm}\begin{tabular}{cc}
\includegraphics[scale=.995]{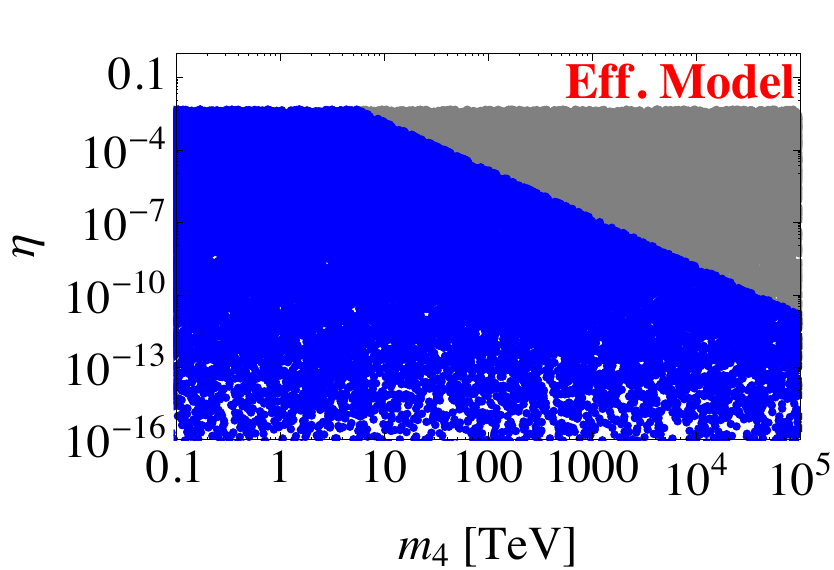}&\includegraphics[scale=.97]{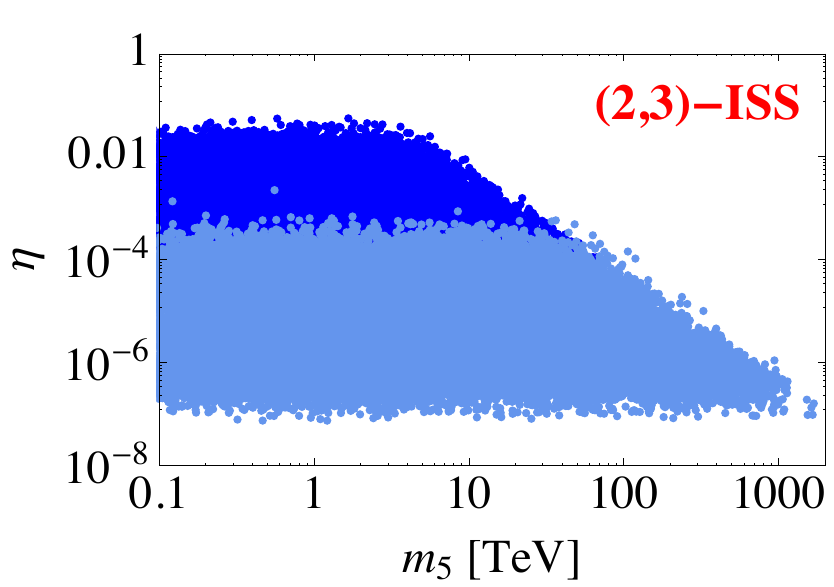} \cr
  \end{tabular}
  \caption{{\footnotesize 
$\eta$ parameter, which  parametrizes the size of mixing between the active and heavy sterile states, is plotted vs the mass of the heavy sterile state. The gray points (left panel) correspond to solutions complying with all experimental data and constraints discussed in the text except for perturbative unitary condition~(\ref{pert-uni}), which we then applied to obtain  the region of dark-blue points. 
In the case of the (2,3)-ISS model (right-panel), 
we  further imposed constraints of Ref.~\cite{Antusch:2014woa} on the matrix $\widetilde \eta$, as well as the bound $B(\mu\rightarrow eee)<10^{-12}$, resulting in the bright-blue region of points. 
 }}
\end{center}
\label{fig:4}
\end{figure}

In Fig.~\ref{fig:4} (left panel), we plot the dependence of $\eta$ with respect to the  mass of the effective sterile neutrino $m_4$. 
Gray points in that plot are obtained by varying the mass of the lightest neutrino, $m_{\nu_e}\in (10^{-21}, 1)$~eV, and by 
imposing the following constraints: (i) Neutrino data (masses and mixing angles) respect the normal hierarchy, with $\Delta m_{21}^2 = 7.5(2) \times 10^{-5}$~eV, and $\Delta m_{31}^2 = 2.46(5) \times 10^{-3}$~eV~\cite{Gonzalez-Garcia:2014bfa}. We checked to see that our 
final results do not change in any significant manner if the inverse hierarchy is adopted. Furthermore, we vary the three mixing angles with the fourth neutrino by assuming $\theta_{i4}\in (0, 2\pi ]$, while keeping the other three mixing angles to 
their best-fit values, namely $\sin^2 \theta_{12} =0.30(1)$,  $\sin^2 \theta_{23} =0.47(4)$,  $\sin^2 \theta_{13} =0.022(1)$~\cite{Gonzalez-Garcia:2014bfa}. (ii) The selected points satisfy the upper bound ${\rm B}(\mu \to e\gamma)< 5.7 \times 10^{-13}$~\cite{Adam:2013mnn}. (iii) The results for 
$R_\pi = \Gamma(\pi \to e \bar \nu_e)/ \Gamma(\pi \to \mu \bar \nu_\mu )$, $R_K$, $\Gamma(W\to \ell\nu_\ell)$, and $\Gamma(Z\to \text{invisible})$, remain consistent with experimental findings. 
We see that for all (heavy) sterile neutrino masses the unitarity breaking parameter is $\eta \lesssim 0.005$. 
That parameter space is not compatible with the perturbative unitarity requirement, which for $m_{4}\gg m_W$ translates into~\cite{Ilakovac:1994kj},~\footnote{To write it in the form given in Eq.~(\ref{pert-uni}), we replaced $\alpha_W=g^2/(4\pi) = \sqrt{2} G_F m_W^2/\pi$. }
\bea\label{pert-uni}
{G_F m_4^2 \over \sqrt{2}\pi}\sum_\alpha \vert U_{\alpha 4}\vert^2  < 1\,.
\eea
The resulting region, i.e. the one that satisfies constraints (i), (ii), (iii) and Eq.~(\ref{pert-uni}), is depicted by blue points (the dark region) in Fig.~\ref{fig:4}, where we see that  the parameter $\eta$ is indeed diminishing with the increase of the heavy sterile mass $m_4$. In other words, the decoupling of a very heavy sterile neutrino entails the unitarity of the $3\times 3$ submatrix $\widetilde U_\text{PMNS}$. Decoupling from active neutrinos for very large masses was also explicitly emphasized in Ref.~\cite{Alonso:2012ji}.
We should mention that, besides the above constraints, we also implemented the constraint coming from ${\rm B}(\mu \to eee) < 10^{-12}$~\cite{Bellgardt:1987du}, but it turns out that the present experimental bound does not bring any additional improvement.  

By imposing the constraints (i) and Eq.~(\ref{pert-uni}) on the (2,3)-ISS model, we get a similar region of allowed (blue) points in the right panel of Fig.~\ref{fig:4}. A notable difference with respect to the situation with one effective sterile neutrino is that the region of very small mixing angles is excluded due to relations between the active neutrino masses and the active-sterile neutrino mixing, cf. Ref.~\cite{Abada:2014vea}. For very heavy $m_5$, on the other hand, the range of allowed $\eta$'s shrinks and eventually vanishes with $m_5\to \infty$.~\footnote{We recall that, in the (2,3)-ISS model, $m_4$ stands for the mass of the light sterile state whose impact on the decays discussed here is negligible [as seen from Eq.~(\ref{eq:Wils})], while $m_5 > m_4$ can be large and is important for B($V\to \ell_\alpha\ell_\beta$).} 
Furthermore, we use the results of Ref.~\cite{Antusch:2014woa} which are derived in the minimal unitarity violation scheme in which the heavy sterile neutrino fields are integrated out, and therefore the observables computed in that scheme are functions of the deviation of PMNS matrix from unitarity only~\cite{MUV}. We adapt and apply them to our (2,3)-ISS model and get a region of the bright-blue points, as shown in the right panel of Fig.~\ref{fig:4}. 
To further constrain the parameter space we find it useful to account for the experimental bound on ${\rm B}(\mu \to eee)<1\times 10^{-12}$, as is discussed in Refs.~\cite{Ilakovac:1994kj,mueee,Alonso:2012ji}.
This latter constraint appears to be superfluous in most of the parameter space, once the constraints of Eq.~(\ref{pert-uni}) and Ref.~\cite{Antusch:2014woa} are taken into account, except in the range $10\ \mathrm{TeV}\lesssim m_5 \lesssim 100\ \mathrm{TeV}$, where the bound ${\rm B}(\mu \to eee)<1\times 10^{-12}$ restricts the parameter space relevant to B($V \to e \mu$).

We also mention that we attempted implementing the constraints coming from various laboratory experiments, summarized in Ref.~\cite{Atre:2009rg}, but since those results only impact the region of relatively small sterile neutrino masses ($m_5 \lesssim 100$ GeV), they are of no relevance to the present study. 
\begin{center}
\begin{figure}
\hspace*{-11mm}\begin{tabular}{cc}
\includegraphics[scale=.45]{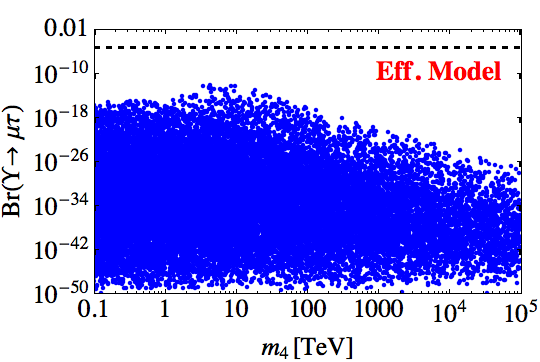}&\includegraphics[scale=.45]{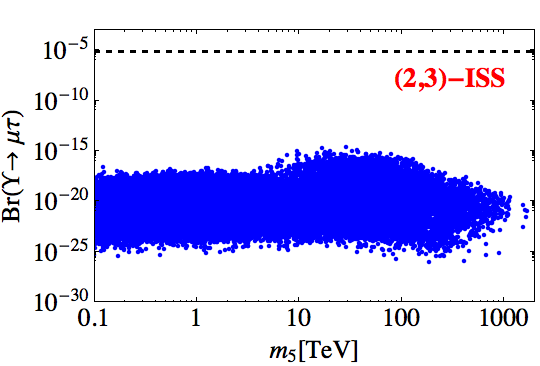} \cr
\includegraphics[scale=.45]{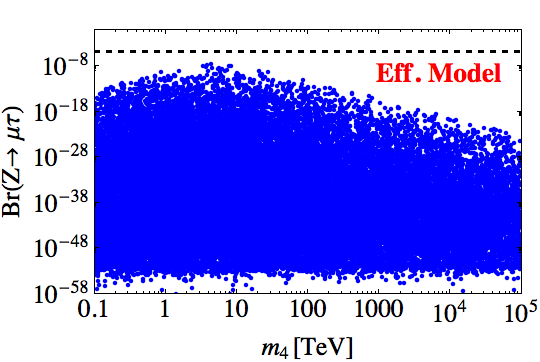}&\includegraphics[scale=.45]{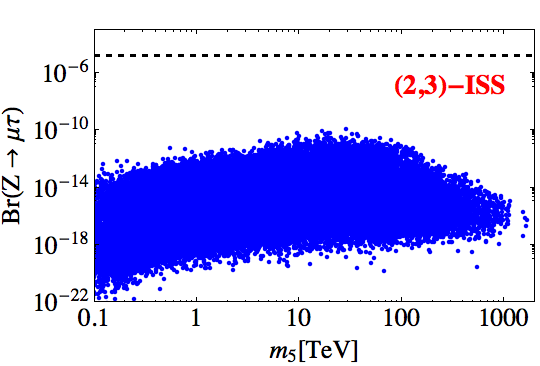} \cr
  \end{tabular}
  \caption{{\footnotesize 
${\rm B}(\Upsilon \to \mu \tau)$ and ${\rm B}(Z \to \mu \tau)$ are shown as functions of the heavy sterile neutrino(s) mass, and in both models considered in this paper. 
The points are selected in such a way that the models are consistent with the constraints discussed in the text and shown in Fig.~\ref{fig:4}.
Dashed horizontal lines correspond to the current experimental upper bounds for these decay rates. Notice again that the mass of the heavy sterile neutrino is denoted as $m_4$ in the effective model, and $m_5$ in the (2,3)-ISS model because the latter contains a lighter sterile neutrino state, the impact of which  is negligible on the decay modes discussed here. 
 }}
\label{fig:5}
\end{figure}
\end{center}

\begin{table}[ht!]
\renewcommand{\arraystretch}{1.5}
\centering{}%
\resizebox{\textwidth}{!}{
\begin{tabular}{|cc|ccc|ccc|}
\hline 
$V$ & $\ell_\alpha \ell_\beta$ & $m_4 = 1$~TeV & $10$~TeV & $100$~TeV& $m_5 = 1$~TeV & $10$~TeV & $100$~TeV \\ \hline\hline
$\phi$ & $e\mu$  & $1 \times 10^{-24}$ & $5 \times 10^{-24}$ & $3 \times 10^{-24}$ & $1\times 10^{-23}$ & $6 \times 10^{-23}$ & $5 \times 10^{-23}$ \\ \hline
$J/\psi$ & $e\mu$  & $2 \times 10^{-21}$ & $3 \times 10^{-20}$ & $6 \times 10^{-21}$ & $2\times 10^{-20}$ & $9 \times 10^{-20}$ & $7 \times 10^{-20}$ \\  
             & $e\tau$  & $5 \times 10^{-18}$ & $8\times 10^{-17}$ & $2\times 10^{-19}$ & $1 \times 10^{-19}$ & $3\times 10^{-18}$ & $1 \times 10^{-19}$ \\  
             & $\mu\tau$  & $8 \times 10^{-18}$ & $6 \times 10^{-16}$ & $3 \times 10^{-20}$ & $4 \times 10^{-19}$ & $4 \times 10^{-18}$ & $8 \times 10^{-19}$ \\ \hline
$\psi (2S)$ & $e\mu$  & $9 \times 10^{-22}$ & $ 1.5 \times 10^{-20}$ & $3 \times 10^{-21}$ & $4 \times 10^{-21}$& $3\times 10^{-20}$ & $2 \times 10^{-20}$ \\  
             & $e\tau$  & $5 \times 10^{-18}$ & $2 \times 10^{-17}$ & $9 \times 10^{-21}$ & $4\times 10^{-20}$ & $1\times 10^{-18}$ & $4 \times 10^{-20}$ \\  
             & $\mu\tau$  & $8 \times 10^{-18}$ & $3 \times 10^{-17}$ & $1.2 \times 10^{-20}$ & $1 \times 10^{-19}$ & $1\times 10^{-18}$ & $2\times 10^{-19}$ \\ \hline
$\Upsilon$ & $e\mu$  & $7 \times 10^{-18}$ & $2 \times 10^{-17}$ & $6 \times 10^{-18}$ & $2 \times 10^{-19}$ & $2\times 10^{-17}$ & $2\times 10^{-17}$ \\  
             & $e\tau$  & $5 \times 10^{-14}$ & $2 \times 10^{-13}$ & $9 \times 10^{-17}$ & $6\times 10^{-18}$ & $4\times 10^{-16}$ & $5\times 10^{-17}$ \\  
             & $\mu\tau$  & $5\times 10^{-16}$  & $2.5\times 10^{-13}$ & $1.2\times 10^{-16}$ & $1\times 10^{-17}$& $8\times 10^{-16}$ &$3\times 10^{-16}$ \\ \hline
$\Upsilon(2S)$ & $e\mu$  & $5\times 10^{-18}$  & $5\times 10^{-18}$ & $1.5 \times 10^{-18}$ & $2 \times 10^{-19}$ & $2\times 10^{-17}$ & $2\times 10^{-17}$ \\  
             & $e\tau$  & $1.8 \times 10^{-14}$ & $3 \times 10^{-14}$ & $3\times 10^{-18}$ & $8\times 10^{-18}$ & $5 \times 10^{-16}$ & $5\times 10^{-17}$ \\  
             & $\mu\tau$  & $2 \times 10^{-16}$ & $2 \times 10^{-13}$ & $2 \times 10^{-17}$ & $2\times 10^{-17}$ & $8 \times 10^{-16}$ & $3 \times 10^{-16}$ \\ \hline
$\Upsilon(3S)$ & $e\mu$  & $1.5 \times 10^{-17}$ & $3 \times 10^{-17}$ & $1.5 \times 10^{-17}$ & $5 \times 10^{-19}$ & $5 \times 10^{-17}$ & $4\times 10^{-17}$ \\  
             & $e\tau$  & $5.5 \times 10^{-14}$ & $3 \times 10^{-14}$ & $4 \times 10^{-17}$ & $2\times 10^{-17}$ & $1\times 10^{-15}$ & $1 \times 10^{-16}$ \\  
             & $\mu\tau$  & $2 \times 10^{-15}$ & $2 \times 10^{-12}$ & $4 \times 10^{-17}$ &  $3 \times 10^{-17}$ & $2 \times 10^{-15}$ & $6\times 10^{-16}$ \\ \hline
$Z$ & $e\mu$  & $1.2\times 10^{-14}$ & $7\times 10^{-13}$ & $4 \times 10^{-13}$& $9\times 10^{-14}$& $8 \times 10^{-13}$ & $6 \times 10^{-13}$ \\  
             & $e\tau$  & $2\times 10^{-10}$ & $9 \times 10^{-9}$ & $4 \times 10^{-13}$ & $7\times 10^{-13}$ & $4\times 10^{-11}$ & $2 \times 10^{-12}$ \\  
             & $\mu\tau$  & $5.5\times 10^{-10}$  & $3.5\times 10^{-8}$ & $1.6\times 10^{-12}$ & $3\times 10^{-12}$& $6\times 10^{-11}$ &$1\times 10^{-11}$ \\ \hline
\end{tabular}
}
\caption{{\footnotesize{}\label{tab:results} Upper bound on ${\rm B}(V\to \ell_\alpha \ell_\beta)$ for three values of the mass $m_{4,5}$. The numbers in the three columns referring to $m_4$ are obtained by using the effective model discussed in the text, while the other three, referring to $m_5$, are results of the (2,3)-ISS model (also discussed in the text). }}
\end{table}
After having completed the discussion on several constraints, we present our results for branching fractions B($V \to \mu  \tau$) depending on the mass of heavy sterile neutrino(s). In Fig.~\ref{fig:5} 
we plot our results for $V=\Upsilon$ and $V=Z$, for which the enhancement is more pronounced. Other cases of $V$ result in similar shapes but the upper bound becomes lower. 
In Table~\ref{tab:results} we collect our results for three values of the heavy sterile neutrino(s) mass. 

To better appreciate the enhancement of the LFV decay rates shown in Fig.~\ref{fig:5}, we emphasize that both of them are ${\rm B}(V \to \mu  \tau) < 10^{-50}$ in the absence of heavy sterile neutrinos. 
Current experimental bounds in both cases are shown by dashed lines. Since those bounds are expected to improve in the near future, a possibility of seeing the LFV modes discussed in this paper might become realistic. 
Conversely, an observation of the LFV modes $V \to \ell_\alpha  \ell_\beta$, with branching fractions significantly larger than the bounds presented in Table~\ref{tab:results} would be a way to disfavor many of
the models containing heavy sterile neutrinos as being the unique source of lepton flavor violation. In obtaining the bounds presented in Table~\ref{tab:results} we used masses and decay constants listed in Appendix~B. 
In presenting our results (the upper bounds) for lepton flavor violating modes, we used the parameters from Ref.~\cite{Antusch:2014woa} which were determined at  90\% C.L. 
For that reason, we treated all  other input data to  2 $\sigma$ as well. Therefore, our final results  in Table~\ref{tab:results} are also obtained at 2 $\sigma$ level.

Finally, we compare in Table~\ref{tab:compare} our upper bounds for the modes for which we could find predictions in the literature.
\begin{table}[h!!]
\renewcommand{\arraystretch}{1.5}
\centering{}%
\hspace*{-4mm}\begin{tabular}{|cccccc|}
\hline 
Mode & Ref.~\cite{Nussinov:2000nm} & Ref.~\cite{Gutsche:2009vp} &  Ref.~\cite{Sun:2012yq}  & Eff. model & (2,3)-ISS \\ \hline 
${\rm B}(\phi \to e\mu)$ & $<4\times 10^{-17}$  & $<1.3 \times 10^{-21}$ & $<5\times 10^{-20}$ & $< 5\times 10^{-24}$ & $<6\times 10^{-23}$ \\  
${\rm B}(J/\psi \to e\mu)$ & $<4\times 10^{-13}$  & $<3.5 \times 10^{-13}$ & $<1.9\times 10^{-18}$ & $< 3\times  10^{-20}$ & $< 9\times 10^{-20}$ \\
${\rm B}(J/\psi \to \mu \tau )$ & $-$ & $-$ & $< 1.6 \times 10^{-7}$ & $< 6\times  10^{-16}$  & $< 4\times 10^{-18}$ \\ 
${\rm B}(\Upsilon \to e\mu)$ & $< 2\times 10^{-9}$  & $<3.8 \times 10^{-6}$ & $< 3.6\times 10^{-18}$ & $< 2\times  10^{-17}$ & $< 2\times 10^{-17}$ \\
${\rm B}(\Upsilon \to \mu \tau )$ & $-$ & $-$ & $< 5.3 \times 10^{-7}$ & $< 2.5\times  10^{-13}$  & $< 8\times 10^{-16}$ \\ 
${\rm B}(Z \to e\mu)$ & $< 5\times 10^{-13}$  & $< 8 \times 10^{-15}$ & $-$ & $< 7\times  10^{-13}$ & $< 8\times 10^{-13}$  \\ \hline
\end{tabular}\caption{{\footnotesize{}\label{tab:compare} Upper bounds ${\rm B}(V\to \ell_\alpha \ell_\beta)$: Comparison of the results reported in the literature with the bounds 
obtained in this work by using two different models (the last two columns). The bounds for other similar decay modes that have not been discussed in the literature can be found in 
Table~\ref{tab:results}. }}
\end{table}

\section{Conclusions}\label{sec:concl}

In this paper we discussed the enhancement of the LFV decays of flavorless vector bosons,  $V \to \ell_\alpha  \ell_\beta$, with $V\in \left\{\phi, \psi^{(n)},\Upsilon^{(n)},Z\right\}$, induced by a mixing between the active and sterile neutrinos. 
The enhancement grows with the mass of the heavy sterile neutrino(s), as can be seen from the mass dependence of the Wilson coefficients that we explicitly calculated. We find that the most significant diagram 
that gives rise to the LFV decay amplitudes is the one coming from the $Z\nu \nu$ vertex, which suggests a steady growth of the decay rate with the mass of the sterile neutrino(s). In the physical amplitude, however, the region of very large 
mass of the sterile neutrino(s) is suppressed as the decoupling takes place, i.e. mixing between the active and sterile neutrinos rapidly falls. 

We illustrated the enhancement of ${\rm B}(V \to \ell_\alpha  \ell_\beta)$ in two scenarios: a model with one effective sterile neutrino that mimics the effect of a generic extensions of the SM including heavy sterile fermions, 
and in a minimal realization of the inverse seesaw scenario compatible with current observations. 
Our results for upper bounds on  ${\rm B}(V \to \ell_\alpha  \ell_\beta)$  [$V\in \left\{\phi,J/\psi, \psi(2S),\Upsilon(1S),\Upsilon(2S),\Upsilon(3S),Z\right\}$] are still considerably smaller than the current experimental bounds (when available), 
but that situation might change in the future as more experimental research will be conducted at Belle II, BESIII, LHC, and hopefully at FCC-ee (TLEP). 
If one of the decays studied here is observed and turns out to have a branching fraction larger than the upper bounds reported here, then sources of LFV other than those coming from mixing with heavy sterile neutrinos must be accounted for.

\section*{Acknowledgements}
We gratefully acknowledge a partial support from the European Union, FP7 ITN INVISIBLES (Marie Curie Actions, PITN-\-GA-\-2011-\-289442). M.L. thanks M.B. Gavela for the interesting comments and discussions.

\newpage 
\section*{Appendix A: Wilson Coefficients} 

\begin{center}
\begin{figure}
\begin{tabular}{cc}
\includegraphics[scale=.44]{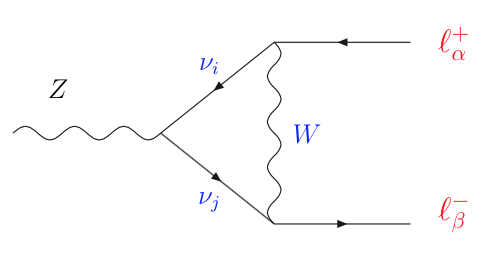}&\includegraphics[scale=.44]{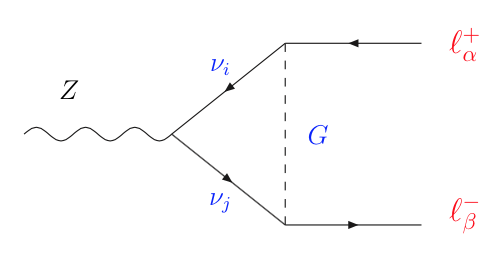} \cr
\includegraphics[scale=.44]{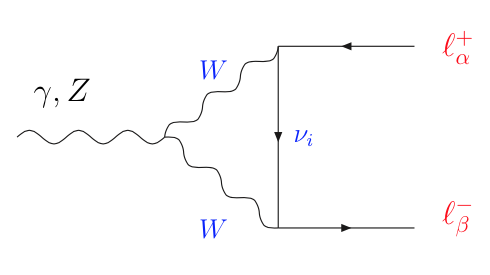}&\includegraphics[scale=.44]{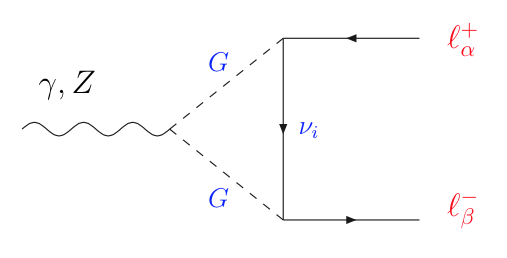} \cr
\includegraphics[scale=.44]{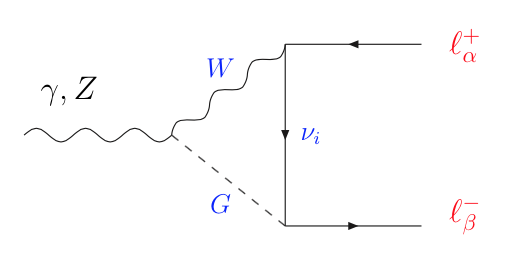}&\includegraphics[scale=.44]{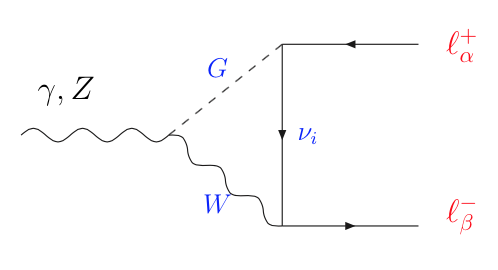} \cr
\includegraphics[scale=.44]{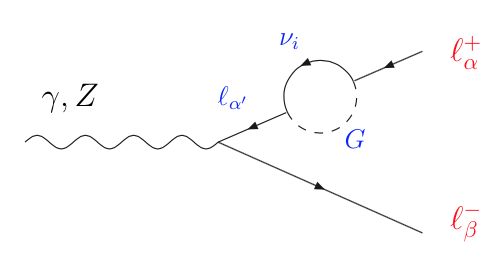}&\includegraphics[scale=.44]{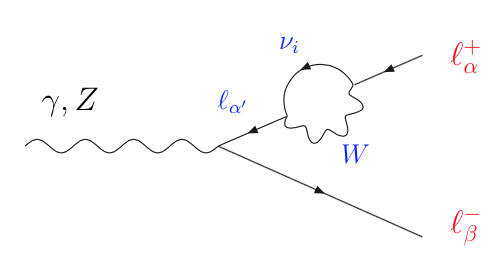} \cr
\includegraphics[scale=.44]{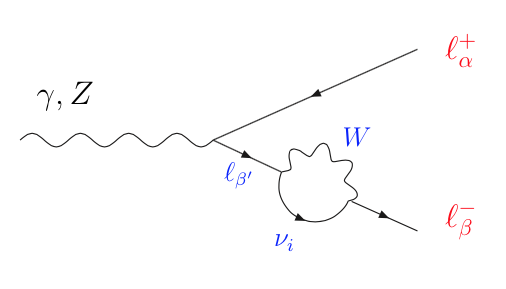}&\includegraphics[scale=.44]{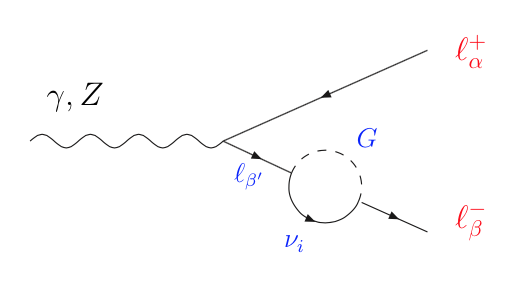} \cr
  \end{tabular}
  \caption{{\footnotesize 
Penguin and self-energy diagrams contributing the LFV decay in Feynman gauge. 
 }}
\label{fig:AA}
\end{figure}
\end{center}

\begin{figure}
\begin{tabular}{cc}
\includegraphics[scale=.4]{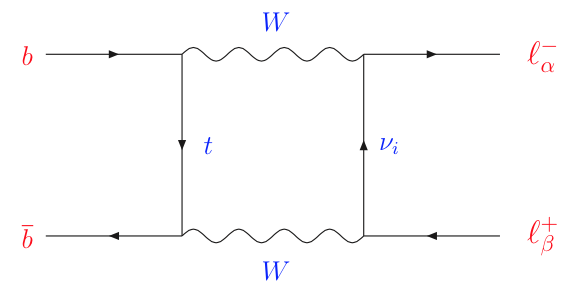}&\includegraphics[scale=.4]{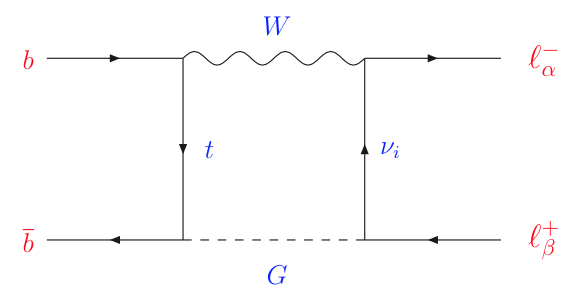} \cr
\includegraphics[scale=.4]{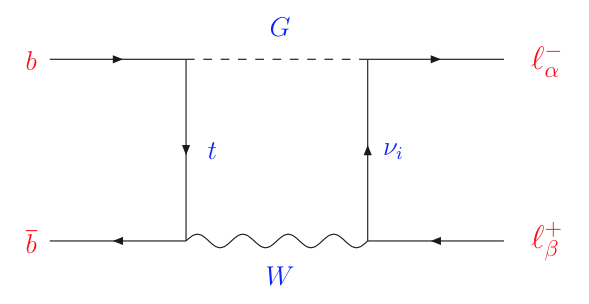}&\includegraphics[scale=.4]{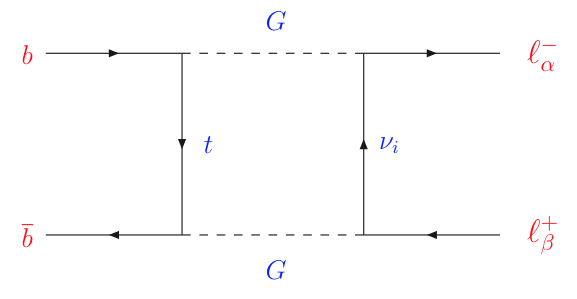} \cr
  \end{tabular}
  \caption{{\footnotesize 
Box diagrams contributing the LFV decay $\Upsilon^{(n)}\to \ell_\alpha\ell_\beta$ in Feynman gauge. 
 }}
\label{fig:B}
\end{figure}

In this Appendix we present detailed expressions for the Wilson coefficients.  
All computations have been made in the Feynman gauge. 
Contributions coming from the penguin and self-energy diagrams are shown in Fig.~\ref{fig:AA}, whereas the box diagrams are shown in Fig.~\ref{fig:B}. 

We use the standard notation, $x_i=m_i^2/m_W^2$, $x_t=m_t^2/m_W^2$, $x_q=q^2/m_W^2= m_V^2/m_W^2$, and write 
\begin{equation}
C_{VL}^r=\displaystyle\sum_{i,j=1}^{n_\nu} U_{\beta i} U_{\alpha j}^* C^{r, ij}_{VL} (x_i,x_j),
\end{equation}
where $r\in\lbrace\gamma,Z,\text{box}\rbrace$. The coefficients $C^{r, ij}_{VL}$ related to $\gamma$ and the box contributions are diagonal, $C^{r, ij}_{VL}= \delta_{ij}C^{r, i}_{VL}$, while 
those related to the $Z$ penguins can also involve a coupling to two different neutrinos, since the $3\times 3$ mixing matrix is no longer unitary. 
We therefore separate the diagonal and nondiagonal parts of the corresponding coefficient $C^{Z, ij}_{VL}=\delta_{ij} C^{Z,i}+\widehat{C}^{Z,ij}$, where the second term depends on the parameter $C_{ij}$ defined by
\begin{align}
C_{ij} = \sum_{\alpha={e,\mu,\tau}} U_{\alpha i}^* U_{\alpha j},
\end{align}
which, in the presence of sterile neutrinos, is generally different from $\delta_{ij}$.
Furthermore, from the plots presented in the body of the present paper we see that the region of $m_{4,5}\gg m_V$ is particularly interesting because there occurs the enhancement 
of the LFV decay rate. For the sake of clarity we thus expand our expressions in $x_q$ and present here only the dominant terms. We also neglected, in the denominators of the loop integrals, the external momenta since they are 
negligible with respect to heavy neutrino masses.
Therefore, up to terms ${\cal O}(x_q^2)$, our results read:
\begin{align}
\label{cvgamma}
	C_{VL}^{\gamma,i} (x_i) = -\frac{1}{16\pi^2} + x_q\frac{-43 x_i^3 + 108 x_i^2 +6 (5 x_i-6)x_i^2 \log x_i-81 x_i+16}{288 \pi^2 (x_i-1)^4} ,
\end{align}

\begin{align}
	C_{VL}^{Z,i}(x_i) =  &  \frac{-1+12 x_i-11 x_i^2+10 x_i^2 \log x_i }{64 \pi^2 (x_i-1)^2} +   \cos^2 \theta_W   C_{VL}^{\gamma,i} (x_i)  ,
\end{align}

\begin{align}
	\widehat{C}_{VL}^{Z,ii}&(x_i,x_i) =C_{ii} \frac{(x_i-2) \left(3 \left(x_i^2-1\right)+2
   (x_i-4) x_i \log x_i\right)}{128 \pi^2 (x_i-1)^2}\\
	&-x_q C_{ii}\frac{(x_i-1) (x_i (x_i (2 x_i-47)+25)+14)+6 (x_i (12 x_i-13)+2) \log x_i}{1152 \pi^2 (x_i-1)^4},\nonumber
\end{align}

\begin{align}
	\widehat{C}_{VL}^{Z,ij}(x_i,x_j)&= \frac{\sqrt{x_i x_j} C_{ij}^*}{64 \pi^2} \Big{[}\frac{x_i (x_i-4)}{(x_i-1)(x_i-x_j)}\log x_i+\frac{x_j (x_j-4)}{(x_j-1)(x_j-x_i)}\log x_j-\frac{3}{2}\Big{]} \\
	&+\frac{C_{ij}}{64 \pi^2}\Big{[} \frac{2 x_i^2 (x_j-1)}{(x_i-1)(x_i-x_j)}\log x_i+\frac{2 x_j^2 (x_i-1)}{(x_j-1)(x_j-x_i)}\log x_j+3\Big{]}\nonumber \\
	&+\frac{x_q}{192 \pi^2} \Big{\lbrace} \sqrt{x_i x_j} C_{ij}^*\Big{[} \frac{x_i^2(x_i-3 x_j+2 x_i x_j)}{(x_i-1)^2(x_i-x_j)^3}\log x_i+\frac{x_j^2(x_j-3 x_i+2 x_i x_j)}{(x_j-1)^2(x_j-x_i)^3}\log x_j\nonumber\\
	&-\frac{x_i^3(x_j-1)-x_j^3(x_i-1)+x_i x_j (x_i-x_j)}{(x_i-1)(x_j-1)(x_i-x_j)^3}\Big{]}+ 2 C_{ij} \Big{[} \frac{x_i^2(3 x_i^2 +3x_j^2+3x_j-x_i-8x_i x_j)}{(x_i-1)^2(x_i-x_j)^3}\log x_i \nonumber \\
	&+\frac{x_j^2(3 x_j^2 +3x_i^2+3x_i-x_j-8x_i x_j)}{(x_j-1)^2(x_j-x_i)^3}\log x_j - \frac{8 x_i^2 x_j-8 x_i x_j^2-x_i^3 x_j+x_i x_j^3-2x_i^3+2x_j^3}{(x_i-1)(x_j-1)(x_i-x_j)^3}\Big{]}\Big{\rbrace},\nonumber
\end{align}

\begin{align}
\label{cvbox}
	C_{VL}^{\mathrm{Box},i} = \frac{1}{256 \pi^2} 
	 \left\lbrace \frac{[ x_i(x_t-8)+4]x_t^2 \log x_t}{(x_t-1)^2(x_i-x_t)}+\frac{[ x_t(x_i-8)+4]x_i^2 \log x_i}{(x_i-1)^2 (x_t-x_i)}+\frac{7 x_i x_t-4}{(x_i-1)(x_t-1)}\right\rbrace.
\end{align}

\section*{Appendix B: Formulas and hadronic quantities} 

In this Appendix we collect the expressions used to constrain the parameters of the models discussed in the present paper, as well as the values of the masses and decay constants used in our numerical analysis.
In the expressions below we used the value of $G_F=G_\mu = 1.166 \times 10^{-5} \ \gev^{-2}$, as extracted from $\mu \to e\nu_\mu\bar\nu_e$. 
In our scenarios, in which we extended the neutrino sector by adding heavy sterile neutrinos, the Fermi constant becomes $G_F=G_\mu/\sqrt{ \sum_{i,j} |U_{ei}|^2  |U_{\mu j}|^2}$. For the models used in this paper, we checked to see that 
$G_F = G_\mu$ remains an excellent approximation.  
 
\begin{itemize}
\item \underline{$\mu \to e\gamma$}: We use the experimentally established upper bound ${\rm B}(\mu\to e\gamma) < 5.7\times 10^{-13}$, and the expression~\cite{Ilakovac:1994kj}
\begin{align}
{\rm B}(\mu \to e\gamma) &= {\sqrt{2} G_F^3 s_W^2 m_W^2 \over 128 \pi^5 \Gamma_\mu} m_\mu^5 \vert U_{\mu 4}^\ast U_{e4} G_\gamma (x_4)\vert^2\,, \cr
G_\gamma (x) & = - { 2 x^3+5x^2 -x\over 4(1-x)^3 }
- {  3 x^3  \over 2(1-x)^4 } \log x\ ,
\end{align} 
to get one of the most significant constraints in this study. Notice that we use $s_W^2=1-m_W^2/m_Z^2$, and we kept the dominant contribution with $x_4$. 
\item \underline{$W \to \ell_\alpha \nu$}: Combining the measured ${\rm B}(W\to e\nu )=0.1071(16)$ and ${\rm B}(W\to \mu\nu)=0.1063(15)$, with the expression
\begin{align}
{\rm B}(W \to \ell_\alpha \nu) = {\sqrt{2} G_F m_W \over 24 \pi  \Gamma_W} \sum_{j=1}^4 \lambda(m_\alpha^2,m_j^2,m_W^2) \left( 
2 - { m_\alpha^2+m_j^2\over m_W^2}- { (m_\alpha^2-m_j^2)^2\over m_W^4} 
\right) \vert U_{\alpha j}^2\vert  ,
\end{align} 
we further restrain the possible values of $m_4$ while varying the mixing angles in the largest possible range.  
\item  \underline{$\Delta r_{K,\pi}= R_{K,\pi}^{\rm exp.}/R_{K,\pi}^{\rm SM} - 1$}: The ratio of the leptonic decay widths of a given meson $P$, $R_P=\Gamma(P\to e \nu_e )/\Gamma(P\to \mu \nu_\mu )$ was  recently shown to be quite restrictive on 
the possible values of $m_{4,5}$ and $\eta$~\cite{Abada:2013aba}.  The most significant constraints actually come from $\Delta r_{\pi}= 0.004(4)$ and  $\Delta r_{K}= -0.004(3)$, and the corresponding formula reads,
\begin{align}
\Delta r_P= - 1 + {m_\mu^2 (m_P^2-m_\mu^2)^2\over m_e^2 (m_P^2-m_e^2)^2} {\displaystyle{\sum_i} \vert U_{ei}\vert^2 \left[ m_P^2 (m_{\nu_i}^2 + m_e^2) -  (m_{\nu_i}^2 - m_e^2)^2 \right] \lambda^{1/2}(m_P^2,m_{\nu_i}^2 ,m_e^2) 
\over \displaystyle{\sum_i} \vert U_{\mu i}\vert^2 \left[ m_P^2 (m_{\nu_i}^2 + m_\mu^2) -  (m_{\nu_i}^2 - m_\mu^2)^2 \right] \lambda^{1/2}(m_P^2,m_{\nu_i}^2 ,m_\mu^2)
}.
\end{align} 
\item \underline{$Z \to \nu \nu$}: To saturate the experimental ${\Gamma}(Z\to \text{invisible} )=0.499(15)$~GeV, we sum over the kinematically available channels involving active and sterile neutrinos,
\begin{align}
\Gamma(Z \to \nu \nu) =& \sum_{i,j} \left(1-\frac{\delta_{ij}}{2}\right)  {G_F \over 12\sqrt{2} \pi m_Z}\lambda^{1/2}(m_Z^2,m_i^2,m_j^2) |C_{ij}|^2 \nn\\
                                              &\times  \left[  2 m_Z^2-m_i^2-m_j^2 - 6m_im_j - {(m_i^2-m_j^2)^2\over m_Z^2}\right]\,.
\end{align} 
\item \underline{$\mu \to e e e$}: We use the experimental upper bound ${\rm B}(\mu \to e e e) <1 \times 10^{-12}$~\cite{Bellgardt:1987du}, and the expression~\cite{Ilakovac:1994kj}
\bee
{\rm B}(\mu \to eee) &=& \frac{G_F^4 m_W^4 }{6144 \pi^7}\frac{m^5_\mu}{\Gamma_\mu}\non 
&&  \left\{ 2 \left|\frac{1}{2}F^{\mu eee}_{\rm Box}+F^{\mu e}_Z-2\sin^2\theta_W (F^{\mu e}_Z-F^{\mu e}_\gamma)\right|^2+4 \sin^4\theta_W \left|F^{\mu e}_Z-F^{\mu e}_\gamma\right|^2 \right. \non 
&& \left.		+ 16 \sin^2\theta_W \mathrm{Re} \left[	(F^{\mu e}_Z +\frac{1}{2}F^{\mu eee}_{\rm Box})	{G^{\mu e}_\gamma}^* 			\right]		- 48 \sin^4\theta_W \mathrm{Re}\left[	(F^{\mu e}_Z-F^{\mu e}_\gamma)	{G^{\mu e}_\gamma}^* 			\right] \right. \non 
&& \left.
		+32 \sin^4\theta_W |G^{\mu e}_\gamma|^2\left[		\ln \frac{m^2_\mu}{m^2_{e}} -\frac{11}{4}		\right]		\right\},
\eee
with the loop functions $F^{\mu eee}_{\rm Box},F^{\mu e}_Z,F^{\mu e}_\gamma,G^{\mu e}_\gamma$ defined in~\cite{Alonso:2012ji}.

\end{itemize}

Finally, the values of hadronic quantities not discussed in the body of the paper but used in our numerical analysis are listed in Table~\ref{tab:costs}.~\footnote{
Notice that the ratio of decay constants $f_{\psi(2S)}/f_{J/\psi}$ has been obtained from the corresponding (measured) electronic widths and the expression $\Gamma (\psi_n\to e^+e^-)= 16 \pi \alpha_{\rm em}^2 f_{\psi_n}^2/(27 m_{\psi_n}^2) $.
 }

\begin{table}[h!!]
\renewcommand{\arraystretch}{1.5}
\centering{}%
\hspace*{-4mm}\begin{tabular}{|ccc|ccc|}
\hline 
Quantity & Value & Ref. & Quantity & Value & Ref.  \\ \hline 
$m_\phi$ & $1.0195$~GeV  & \cite{PDG} & $f_\phi$ & $241(18)$~MeV &  \cite{hpqcd-2} \\ 
$m_{J/\psi}$ & $3.0969$~GeV  & \cite{PDG} & $f_{J/\psi}$ & $418(9)$~MeV &  \cite{c-latt1} \\ 
$m_{\psi(2S)}$ & $3.6861$~GeV  & \cite{PDG} & $f_{\psi(2S)}/f_{J/\psi}$ & $0.713(16)$ &  \cite{PDG} \\ \hline
$m_{\Upsilon}$ & $9.460$~GeV  & \cite{PDG} & $f_{\Upsilon}$ & $649(31)$~MeV &  \cite{Lepage} \\ 
$m_{\Upsilon(2S)}$ & $10.023$~GeV  & \cite{PDG} & $f_{\Upsilon(2S)}$ & $481(39)$~MeV &  \cite{Lepage} \\ 
$m_{\Upsilon(3S)}$ & $10.355$~GeV  & \cite{PDG} & $f_{\Upsilon(3S)}$ & $539(84)$~MeV &  \cite{RL} \\ \hline 
\end{tabular}\caption{{\footnotesize{}\label{tab:costs} Masses and decay constants used in numerical analysis. }}
\end{table}

\end{document}